\documentclass[letterpaper,pra,twocolumn,showpacs,superscriptaddress]{revtex4}

\usepackage{graphicx}
\usepackage{amsmath}
\usepackage{amssymb}

\include{MLv1.bib}

\begin{document}

\title{Generation of Squeezing in Higher Order Hermite-Gaussian Modes with an Optical Parametric Amplifier}

\author{M.Lassen$^{1,2}$}
\email{mlassen@fysik.dtu.dk}
\noaffiliation{}

\author{V.Delaubert$^{1,3}$}
\noaffiliation{}

\author{C.C.Harb$^{1,4}$}
\noaffiliation{}

\author{P.K.Lam$^{1}$}
\noaffiliation{}

\author{N.Treps$^{2}$}
\noaffiliation{}

\author{H-A.Bachor}
\noaffiliation{}

\affiliation{The Australian National University, ACQOA, Canberra ACT 0200, Australia\\
$^{2}$ Department of Physics, DTU, Building 309, DK-2800 Lyngby,
Denmark\\
$^{3}$ Laboratoire Kastler Brossel, 4 place Jussieu, case 74, Paris
75252 Cedex 05, France\\
$^{4}$ School of
Information Technology and Electrical Engineering, University
College, The University of New South Wales, Canberra, ACT, 2600}

\date{\today}

\begin{abstract}

We demonstrate quantum correlations in the transverse plane of
continuous wave light beams by producing $-4.0~dB$, $-2.6~dB$ and
$-1.5~dB$ of squeezing in the TEM$_{00}$, TEM$_{10}$ and
TEM$_{20}$ Hermite-Gauss modes with an optical parametric
amplifier, respectively. This has potential applications in
quantum information networking, enabling parallel quantum
information processing. We describe the setup for the generation
of squeezing and analyze the effects of various experimental
issues such as mode overlap between pump and seed and nonlinear
losses.
\end{abstract}


\maketitle

\section{Introduction}

There has been a growing interest during recent years in spatial
quantum optical effects, usually called quantum imaging effects
\cite{Lugiato2002,Kolobov1999,Gigan2006}, as the generation of
spatial correlations or spatial squeezing in the transverse plane of
laser beams may open the way for new applications in many different
areas. Among them are biophotonics, laser physics, astronomy and
quantum information. Some pertinent examples are the measurement of
small transverse displacement and tilt of a TEM$_{00}$ laser beam
below the quantum noise limit \cite{Treps2002,Delaubert2006},
detection of weak phase images \cite{Lugiato2002}, quantum
teleportation of optical information \cite{Sokolov2001}, transverse
spatial quantum correlation for transmission of images
\cite{Gigan2006} and noiseless image amplification
\cite{Kolobov1995}. Multiple spatial modes can also provide
advantages in regard to the complexity of quantum information
protocols \cite{Caves1994} and can allow parallel transfer of
quantum information through an optical network. In single photon
optics, this has created considerable interest in the use of modes
with different angular momentum \cite{Zeilinger2000,
Oemrawsingh2004, Langford2004, Barnett2005}. An advantage of
continuous wave light beams is that close to perfect modulation and
detection schemes are available, which is a requirement for the
effective use of squeezed and entangled light in quantum information
protocols.

Several orthonormal basis are available to describe the spatial
properties of laser beams. The most commonly used are the
Hermite-Gauss (H-G) and the Laguerre-Gauss basis \cite{Siegman1986}.
In this contribution will we concentrate on the H-G modes, but a
similar study could be undergone with another set of modes. The
higher order H-G modes are particularly interesting with a cartesian
description of the transverse plane, as they are directly related to
simple spatial properties of Gaussian beams \cite{Kogelnik1964}. The
real and imaginary parts of the TEM$_{10}$ mode represent small
changes in tilt and position of a TEM$_{00}$ beam
\cite{Delaubert2006}, whereas the real and imaginary parts of the
TEM$_{20}$ mode correspond to a small waist-size and waist-position
mismatch \cite{Morrison1994}. Using electro-optic spatial
modulators, the information can be encoded directly into the H-G
modes. The information can then be extracted using homodyne
detections with adapted transverse profiles for the local oscillator
\cite{Bachor2006,Delaubert2006}. In this way we can encode and
detect parallel quantum information in the transverse plane of
continuous wave light beams. In order to apply quantum information
protocols to the transverse plane of continuous wave light beams,
what was missing until now was a reliable and efficient source for
the generation of squeezing in different higher order H-G modes. We
propose here a method to fill this gap.

The present paper is devoted to the experimental study of squeezing
in higher order H-G modes using an optical parametric amplifier
(OPA). In the past OPAs have proven to be efficient sources for the
generation of quadrature squeezed light. Since the first generation
of squeezed light by parametric down-conversion by Wu \emph{et al.}
in 1986 \cite{Wu1986}, squeezed noise power variances of up to -7~dB
\cite{Lam1999}, and squeezing at sideband frequencies down to
sub-kHz frequencies \cite{McKenzie2004} have been demonstrated with
OPAs. However squeezing has so far been limited to TEM$_{00}$ mode
operation. We demonstrate here an experimental technique to produce
squeezed light in different H-G modes, namely the TEM$_{10}$ and
TEM$_{20}$ modes, using an OPA. We report -4.0~dB, -2.6~dB and
-1.5~dB of observed squeezing for the TEM$_{00}$, TEM$_{10}$ and
TEM$_{20}$ H-G modes, respectively. The squeezing is generated in a
back seeded OPA with a bulk lithium niobate (MgO:LiNbO{$_{3}$}) type
I nonlinear crystal, pumped by the second harmonic, from a
continuous wave solid-state monolithic YAG laser.

The paper is organized as follows. In section \ref{OPA}, we
introduce a mode-overlap theory for second order nonlinear
parametrical interaction with higher order H-G modes, in the thin
crystal approximation. Section \ref{ex.setup} describes our
experimental setup and procedure to generate squeezing in different
H-G modes with an OPA. Section \ref{optimizing} address experimental
issues such as the optimization of the parametrical interaction.
Section \ref{sql_meaurement} presents our squeezing measurements.
Finally, we conclude and present an outlook for further work in
section \ref{conclusion}.

\section{Optical parametric amplification and ideal pump mode for TEM$_{n0}$
squeezing}\label{OPA}

An OPA is a second order nonlinear optical device, where three
optical beams are coupled parametrically to each other through the
second order susceptibility $\chi^{(2)}$ of a nonlinear crystal, in
an optical cavity. A pump photon of energy $\hbar\omega_p$ incident
on an OPA down-converts into two photons, signal and idler, of
energy $\hbar\omega_{s}$ and $\hbar\omega_i$ respectively, hence the
name parametrical down-conversion \cite{Boyd1992}. In order to have
a significant effect, the nonlinear interaction must satisfy energy
($\omega_p=\omega_s+\omega_i$) and phase-matching ($k_p=k_s+k_i$)
conservation. In our experiment, the frequency of signal and idler
are degenerate, i.e. $\omega_s=\omega_i$,  and the polarization of
the pump is at $90^\circ$ to the polarization of the signal and
idler, corresponding to a type I phase-matching.

In order to pump the crystal efficiently  for a TEM$_{n0}$ signal
mode, the transverse profile of the pump mode must locally match
with the square of the signal mode. We use $\{u_{i}\}$ as the
notation for the fundamental signal and idler mode basis whose
first mode has a waist of $w_{0}$, and $\{v_{i}\}$ for the second
harmonic pump mode basis whose first mode has a waist of
 $w_{0}/\sqrt{2}$. In the thin crystal
approximation and in the case of a TEM$_{n0}$ signal mode, the
optimal pump profile $\mathcal{E}_{n}(r)$ is defined by:
\begin{eqnarray}\label{SHGmodes}
\mathcal{E}_{n}(r) &=& \sum_{i=1}^{n}\Gamma_{ni}v_{2i}(r).
\end{eqnarray}
where $r=\sqrt{x^2+y^2}$ is the transverse beam coordinate and
$\Gamma_{ni}$ describes the spatial overlap between the squared
signal and the second harmonic pump modes in the transverse plane,
and is given by:
\begin{eqnarray}
\label{overlapInt}
\Gamma_{ni}=\int_{-\infty}^{\infty}\frac{u_{n}^{2}(r)}{\alpha_{n}}v_{2i}(r)dr,
\end{eqnarray}
where $\alpha_{n}$ corresponds to the normalization of the squared
signal, and is defined by the relation
$\alpha_{n}^{2}=\int_{-\infty}^{\infty}u_{n}^{4}(r)dr$. The optimal
pump profile has only even components since the TEM$_{n0}$ signal
squared profile is necessarily even, and its highest component is of
order $2n$. This decomposition is finite because the pre-exponential
polynomial in the expression of $u_{n}^{2}(r)$ is of order $2n$, and
therefore does not project onto higher order modes. The common case
of using a TEM$_{00}$ signal mode yields $\Gamma_{00}=1$ and
corresponds to a perfect spatial overlap as the optimal pump profile
is also a TEM$_{00}$ mode. The presence of several non zero
coefficients implies  that for all cases, except a TEM$_{00}$ pump
mode, the optimal pump profile does not correspond to the signal
intensity distribution. For a TEM$_{10}$ signal mode resonant in the
OPA cavity, the only non zero overlap coefficients calculated from
Eq.~\ref{overlapInt} are given by:
\begin{eqnarray}\label{overlapcoeff10}
\Gamma_{10}&=&0.58 \nonumber\\
\Gamma_{12}&=&0.82.
\end{eqnarray}
As they are the only non zero ones, they fulfill
$\Gamma_{10}^{2}+\Gamma_{12}^{2}=1$. For a TEM$_{20}$ signal mode,
the only pump modes which have non zero overlap in the cavity are
the TEM$_{00}$, TEM$_{20}$ and TEM$_{40}$ modes. The overlap
coefficients are given by:
\begin{eqnarray}\label{overlapcoeff20}
\Gamma_{20}&=&0.47 \nonumber\\
\Gamma_{22}&=&0.44 \nonumber \\
\Gamma_{24}&=&0.77.
\end{eqnarray}
Again, they fulfill
$\Gamma_{20}^{2}+\Gamma_{22}^{2}+\Gamma_{24}^{2}=1$. The presence of
several non zero coefficients accounts for the multi-mode aspect of
the ideal pump mode. Although generating such a complicated mode is
in principle possible by using holograms \cite{Vaziri2002} or
forcing a laser cavity to emit in these mode \cite{Lassen2005,
Schwarz2004}, we choose as a first step, to use a TEM$_{00}$ pump
mode, as discussed in the next section.

Still using the thin crystal approximation with a multi-mode
description of each field as in reference \cite{Schwob1998}, we can
show that pumping the OPA cavity with a non optimal mode results in
an increase of the oscillation threshold. This increase is inversely
proportional to the square of the overlap coefficient between the
chosen pump mode and the optimal pump mode. For instance, when the
signal mode resonant in the OPA is a TEM$_{10}$ mode, pumping with a
TEM$_{00}$ mode increases the threshold by a factor
$1/\Gamma_{10}^{2}\simeq3$. As for a TEM$_{20}$ signal mode, pumping
with a TEM$_{00}$ mode increases the threshold by a factor
$1/\Gamma_{20}^{2}\simeq4.5$. Since the threshold is directly
related to the local intensity in the crystal, it will be further
increased for higher order signal mode operations. Indeed, higher
order mode intensity is more spread out in the transverse plane than
for the TEM$_{00}$ mode. Therefore the best conversion efficiency is
expected to happen for a TEM$_{00}$ mode pump, independently from
any overlap issue. Still in the thin crystal approximation, the
threshold is further increased by a factor of $1.3$, and $1.6$,
respectively for TEM$_{10}$ and TEM$_{20}$ signal modes, relative to
a TEM$_{00}$ signal mode operation. The relative theoretical
thresholds for the three first H-G modes are presented in
Table~\ref{threshold} in section \ref{optimizing}.

We have estimated theoretically the threshold modifications imposed
by the operation of an OPA with higher order transverse modes, and
will use this knowledge in the next sections to interpret our
experimental results.

\section{The experimental setup}\label{ex.setup}

\begin{figure}[htbp]
\begin{center}
\includegraphics[width=8cm]{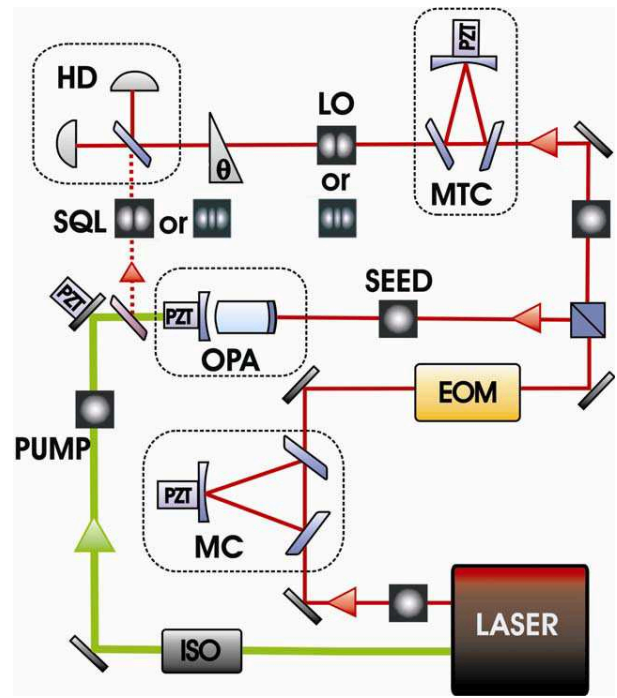}
\caption{ Experimental setup to generate higher order transverse
mode squeezing. An OPA is seeded with a misaligned TEM$_{00}$
beam. The cavity is locked to the fundamental TEM$_{n0}$ mode and
pumped with a second harmonic TEM$_{00}$ beam. The TEM$_{n0}$
squeezed beam is analyzed using a homodyne detection (HD), whose
TEM$_{n0}$ local oscillator is created from a misaligned ring
cavity (MTC).} \label{OPA-Scheme}
\end{center}
\end{figure}

The experimental squeezing setup is illustrated in
Fig.~\ref{OPA-Scheme}. To ensure stable squeezing, 6 locking loops
are implemented in the experiment. For locking we use the Pound
Drever Hall locking technique, where a phase-modulation is
imparted on the optical beams with electric-optic modulators
(EOM). The generated error-signals are then fed back to the
cavities through piezo-electric elements (PZT) \cite{PDH}. All
cavities in the experiment are therefore held at resonance at the
same time. The experiment stays locked for longer than 20 minutes.
The only limiting factor is temperature fluctuations in the
laboratory.

The experimental procedure for producing squeezing is as follows.
The OPA is back seeded with a 1064~nm TEM{$_{00}$} beam. The seed is
misaligned into the OPA in order to excite higher order H-G modes.
The OPA cavity is then locked to the TEM{$_{n0}$} mode and pumped
with a TEM$_{00}$ second harmonic mode, wavelength $\lambda=532$~nm,
generated internal in our cw solid-state monolithic YAG laser. The
laser gives 950~mW of second harmonic power and 195~mW of infrared
power \cite{diabolo}. The 1064~nm beam from the laser is first sent
through a ring cavity, a so-called mode-cleaner (MC), which filters
out the intensity and frequency noise of the laser above the
bandwidth of the MC. The MC also defines a high quality spatial
mode. A bandwidth of 2.5 MHz is measured and a transmission greater
than 90\% is obtained for the TEM$_{00}$ mode. The pump beam from
the internal frequency doubler passes through an optical isolator
(ISO: isolation $>40$~dB) and is carefully mode-matched (95$\%$)
into the OPA. Working at pump powers below threshold, the seed is
either amplified or de-amplified depending on the relative phase
between the pump and the seed. This relative phase is stably locked
to de-amplification in order to generate an amplitude quadrature
squeezed beam.

We used a concentric hemilithic cavity design for the OPA,
consisting of a 2 x 2.5 x 6.5 mm$^3$ lithium niobate
(MgO:LiNbO{$_{3}$}) type I nonlinear crystal doped with 7\% of
magnesium. The birefringence of the crystal is highly temperature
dependent; accurate temperature control of the crystal, typically at
the milli-Kelvin level, is therefore required to achieve optimal
phase-matching, as shown in Fig.~\ref{sinc_fkt}.

The back surface of the crystal is polished so that it has a 8~mm
radius of curvature and is high reflectance coated for both
wavelengths. The output coupler has $96\%$ reflectivity for 1064~nm
and $10\%$ reflectivity for 532~nm, has a radius of curvature of 25
mm, and is placed 23~mm from the front-end of the crystal. The
cavity is therefore near concentricity with a waist of 24~{$\mu$}m
for the 1064~nm cavity and 19 {$\mu$}m for the 532~nm cavity. The
OPA has a finesse of approximately 165 with a free spectral range of
10 GHz and a cold cavity bandwidth of 60 MHz. The TEM{$_{n0}$}
squeezed beam generated at the output of the OPA and the pump beam
are separated with dichroic mirrors. The squeezed beam is then
analyzed using a homodyne detection with a TEM{$_{n0}$} local
oscillator (LO), thus extracting only the information of the
TEM{$_{n0}$} component of the squeezed beam. The LO is created with
a ring cavity, used as a mode transferring cavity (MTC) and designed
to prevent any transverse mode degeneracy when locked to resonance
on the TEM$_{n0}$ mode.

Before describing the measured squeezing we focus on the classical
behavior of the OPA.

\section{Optimizing the parametric interaction: Gain and
threshold measurements}\label{optimizing}

In order to have large parametric interaction we need the best
possible mode overlap between pump and seed modes. To achieve
this, different issues have to be addressed such as mode-matching
and alignment of pump and seed into the cavity. A useful tool for
optimizing the parametric interaction - and at the same time the
possible amount of squeezing to be extracted - is to measure the
classical gain factor of the seed.

Since our OPA is not a cavity for the pump beam, the output coupler
has only 10\% reflectivity for 532~nm, we can envision three
different cases for pumping the OPA efficiently, knowing that we are
restricted to a TEM$_{00}$ pump mode operation, as discussed in the
previous section. We can match the pump profile to the TEM$_{00}$
mode defined through the infrared mode; we can de-focus the pump;
or, we can misalign the pump to match with one lobe (one side) of
the infrared mode, in the TEM$_{10}$ mode case for instance. In each
case we have tuned the crystal temperature to maximize the gain and
find that the most efficient option is to pump with a TEM$_{00}$
mode that is aligned with the cavity axis and optimally mode matched
for maximum coupling to the cavity, as we would do for if we wanted
to produce TEM$_{00}$ squeezing.  The disadvantage of this method
was that the maximum pump power that could be delivered to the
system was set by the threshold power of the TEM$_{00}$.

\begin{figure}[htbp]
\begin{center}
\includegraphics[width=8cm]{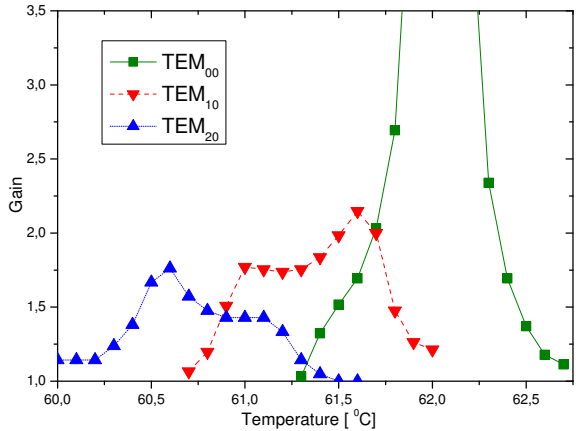}
\caption{Classical gain factor as a function of the crystal
temperature for OPA operation with a TEM$_{00}$, a TEM$_{10}$ and a
TEM$_{20}$ signal modes. The optimal phase-matching temperature are
62.1$^\circ$C, 61.6$^\circ$C and 60.6$^\circ$C for the TEM{$_{00}$},
TEM{$_{10}$} and TEM{$_{20}$} H-G modes, respectively.
 }  \label{sinc_fkt}
\end{center}
\end{figure}

The optimal temperature of the nonlinear crystal to produce
squeezing differs for each TEM$_{n0}$,  due to the different Gouy
phase-shifts between H-G modes \cite{Delaubert2006_2}. The
temperature has thus to be re-optimized for each experiment. A
direct manifestation of the shift in optimized phase-matching
temperature is shown on Fig.~\ref{sinc_fkt}. This figure shows the
classical gain factor measured as a function of the crystal
temperature, for signal profiles given by the three first H-G modes,
still using the best pumping option. We find that the optimal
phase-matching temperature are 62.1$^\circ$C, 61.6$^\circ$C and
60.6$^\circ$C for the TEM{$_{00}$}, TEM{$_{10}$} and TEM{$_{20}$}
modes, respectively. The width (FWHM) for optimal phase-matching
temperature is approximately 1$^\circ$C for all three cases.

\begin{figure}[htbp]
\begin{center}
\includegraphics[width=8.5cm]{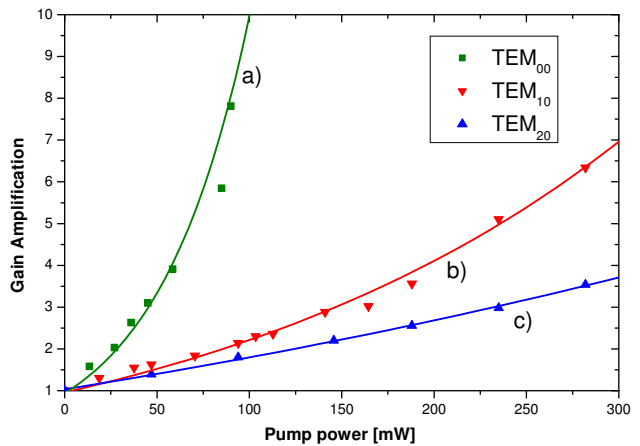}
\caption{Experimental classical amplification gain curves for the
OPA. a) TEM{$_{00}$}, b) TEM{$_{10}$} and c) TEM{$_{20}$} H-G modes.
The solid lines are exponential fits.
 }  \label{amplification}
\end{center}
\end{figure}

The measured amplification gain curves for TEM$_{00}$, TEM$_{10}$
and TEM$_{20}$ H-G modes are presented in Fig.~\ref{amplification}.
We measured a maximum amplification of 300, 23, 5 for the
TEM$_{00}$, TEM$_{10}$ and TEM$_{20}$ H-G modes, respectively. The
best measured de-amplification factors are 0.30, 0.56 and 0.70 for
the TEM$_{00}$, TEM$_{10}$ and TEM$_{20}$ H-G modes, respectively.
On one hand, the de-amplification of 0.30 for the TEM$_{00}$
indicates that the system is close to the oscillation threshold,
since the theoretical value for de-amplification from a back seeded
OPA at threshold is 0.25. On the other hand, the threshold is far
from being reached for the TEM$_{10}$ and TEM$_{20}$ modes cases.
The reason why the de-amplification curves are flat and do not reach
the theoretical value of 0.25 is due to mode-mismatch between the
pump and seed which induces phase noises and losses.

The oscillation threshold is measured to be 260~mW of pump power
when the OPA is resonant for the TEM$_{00}$ signal mode. However,
the threshold for higher order modes cannot be accessed
experimentally because the system starts to oscillate on the
TEM$_{00}$ as soon as the pump power reaches approximately 350~mW,
even when the crystal temperatures are optimized for an operation on
the TEM$_{10}$ and TEM$_{20}$ modes. Nevertheless, we can use the
gain curves obtained experimentally for the TEM$_{10}$ and
TEM$_{20}$ modes with a TEM$_{00}$ pump in order to estimate the
threshold for the TEM$_{10}$ and TEM$_{20}$ modes. A linear fit of
the first couple of data points gives the relative gain slopes
between the TEM$_{10}$, TEM$_{20}$ and TEM$_{00}$ modes, yielding an
estimate of the relative threshold. We find the thresholds for the
TEM$_{10}$ and TEM$_{20}$ OPA operation regimes to be approximately
$1000$~mW and $1600$ mW, respectively. These values can be compared
with a theoretical calculation of the thresholds introduced in
section \ref{OPA}, taking into account the imperfect spatial overlap
between the infrared mode resonant in the cavity and the TEM$_{00}$
pump mode and the lower local intensity for higher order modes. This
comparison is presented in Table~\ref{threshold}, showing a very
good agreement between theory and the experimental measurements.

We have estimated the threshold for each operation of the OPA on
higher order modes, which will allow a calculation of the maximum
amount of squeezing that can be generated by the system in the next
section.
\begin{table}
\centering \label{threshold}
  \begin{tabular}{|l|c|c|c|}
  \hline
        & TEM{$_{00}$} & TEM{$_{10}$} & TEM{$_{20}$} \\ \hline
~Experimental~     & 1 & ~~3.9 $\pm$~0.5~~ & ~~6.2 $\pm$~0.8~~ \\
~Theoretical~~  & 1 & 4 & 7 \\
\hline
\end{tabular}
\caption{Threshold comparison between experimental results and
theory in the thin crystal approximation. The threshold for the
different modes is normalized with respect to the TEM{$_{00}$}
threshold power.} \label{threshold}
\end{table}

\section{ TEM$_{00}$, TEM$_{10}$ and TEM$_{20}$ squeezing }\label{sql_meaurement}

\begin{figure}[htbp]
\begin{center}
\includegraphics[width=8.55cm]{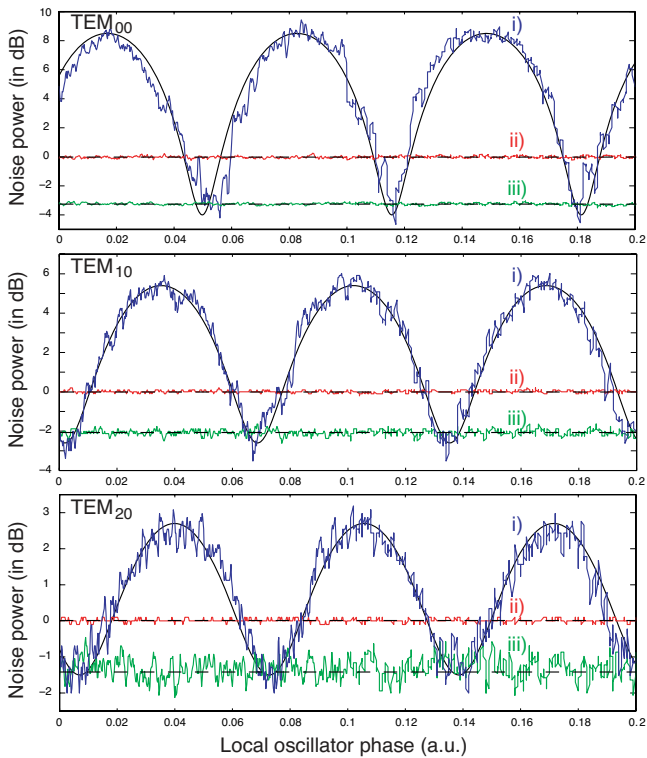}
\caption{Experimental squeezing traces on the a) TEM{$_{00}$}, b)
TEM{$_{10}$} and c) TEM{$_{20}$} modes, recorded by homodyne
detection. i) Scan of the relative phase between the LO and the
squeezed beam. ii) Quantum noise limit. iii) Phase of the LO locked
to the squeezed amplitude quadrature.} \label{SQZ}
\end{center}
\end{figure}
The experimental squeezing curves in the TEM$_{00}$, TEM$_{10}$
and TEM$_{20}$ modes are shown in Fig.~\ref{SQZ}. The squeezing
spectra are recorded on a spectrum analyzer with a resolution
bandwidth of 300 kHz and video bandwidth of 300 Hz at a detection
frequency of 4.5 MHz. All traces are normalized to the quantum
noise level (QNL). The QNL given by trace ii) is measured by
blocking the squeezed beam before the homodyne detector. Trace i)
is obtained by scanning the phase of the LO, and trace iii) by
locking the LO phase to the squeezed amplitude quadrature. The
smooth line is the theoretical fit of the noise variance assuming
the actually experimental parameters. We measured $-4.0\pm0.2$~dB
of squeezing and $+8.5\pm0.5$~dB of anti-squeezing for the
TEM{$_{00}$} mode, $-2.6\pm0.2$~dB of squeezing and
$+5.4\pm0.4$~dB of anti-squeezing for the TEM{$_{10}$} mode, and
$-1.5\pm0.3$~dB of squeezing and $+2.7\pm0.4$~dB of anti-squeezing
for the TEM{$_{20}$} mode. To our knowledge, this is the first
demonstration of higher order transverse mode squeezing using an
OPA. These values have been corrected for electrical noise, which
is $9.1\pm0.1$~dB below the QNL, and is mostly due to the
amplifiers in the photo-detectors.

In order to perform these measurements, we used the following pump
powers which were maximizing the amount of squeezing: $100$~mW for
the TEM$_{00}$ mode and $300$~mW for TEM$_{10}$ and TEM$_{20}$
modes. We were unable to pump the OPA with higher power, as the
system was starting to oscillate on the TEM$_{00}$ mode at 300~mW,
even when the cavity was locked to another mode, as discussed in
the previous section.

We can calculate an upper limit for the measurable noise variance of
the amplitude $V^{-}$ and phase $V^{+}$ quadratures at zero
frequency, using the following simple analytical expression
\cite{BachorBook}:
\begin{eqnarray}\label{SQZeq}
V^{\pm}(0)=1\pm\eta_{esc}\frac{4\sqrt{P/P_{thr}}}{\left(1\mp\sqrt{P/P_{thr}}\right)^{2}},
\end{eqnarray}
where $P$ and $P_{thr}$ correspond to pump and threshold power,
respectively ; $\eta_{esc}= T/(T+L)$ is the escape efficiency, where
$L$ is the intra-cavity loss and $T$ is the transmittance of the
output-coupler. The estimated intra-cavity losses for our OPA is
approximately $L= 0.0043$, where we consider absorption in the
material (0.1\%/cm) and scattering at the mirror and crystal. This
gives a cavity escape efficiency of approximately {$\eta_{cav}$} =
0.89. The calculated squeezing and anti-squeezing for the different
modes, using Eq~\ref{SQZeq} are shown in Table~\ref{sqlamount}c).
The discrepancy between the calculated and experimental values
suggests that there exists losses in our system.

From Fig.~\ref{SQZ}, we can see that the noise spectrum from the OPA
is far away from the minimum uncertainty state predicted for a
lossless OPA. We will in the following estimate and characterize the
losses in our experiment.

The total detection efficiency of our experiment, is given by:
$\eta_{total}=\eta_{cav}\eta_{prop}\eta_{det}\eta_{hd}$, where
$\eta_{prop}=0.97\pm0.02$ is the propagation efficiency,
$\eta_{det}=0.93\pm0.05$ is the photo-detector (Epitaxx ETX500)
efficiency. $\eta_{hd}$ is the homodyne detection efficiency.  By
carefully optimizing the homodyne detection efficiency we measured
$\eta_{hd}(TEM_{00})=0.98\pm 0.02$, $\eta_{hd}(TEM_{10})=0.95\pm
0.02$ and $\eta_{hd}(TEM_{20})=0.91\pm 0.02$ for the different
modes. The mode dependance is firstly due to the larger spatial
extension of higher order modes, which are more apertured by the
optics. Moreover, the fringe visibility drops for higher order modes
because of the additional transverse degree of freedom compared to
the TEM$_{00}$ case. Finally, a small mode mismatch has more
dramatic effects on the fringe visibility for a complex intensity
distribution. The total estimated detection efficiencies for our
experiment are therefore $\eta_{total}=0.79\pm0.04$,
$\eta_{total}=0.76\pm0.04$ and $\eta_{total}=0.73\pm0.04$ for the
TEM{$_{00}$}, TEM{$_{10}$} and TEM{$_{20}$}, respectively. From
these efficiencies can we infer squeezing and anti-squeezing values,
see Table~\ref{sqlamount}. In order to compare these efficiencies,
we can calculate the total theoretical detection efficiency from the
squeezing spectrum. The theoretical detection efficiency for a lossy
OPA is given by:
\begin{eqnarray}\label{lossASQZ}
\eta = \frac{V_{SQL} + V_{ASQL} -1 -V_{SQL}V_{ASQL}}{V_{SQL} +
V_{ASQL} -2},
\end{eqnarray}
where $V_{SQL}$ is the measured squeezing, $V_{ASQL}$ is the
measured anti-squeezing and $\eta$ is the total efficiency of the
experiment. The calculated efficiencies can be seen in
Table~\ref{tablelos}.

\begin{table}
\centering
  \begin{tabular}{|c|c|c|c|c|c|c|}
  \hline
  & \multicolumn{2}{c|}{TEM{$_{00}$}}& \multicolumn{2}{c|}{TEM{$_{10}$}}&
  \multicolumn{2}{c|}{TEM{$_{20}$}}\\ \hline
~~a) Corrected~~       & ~-4.0~ & ~+8.5~ & ~-2.6~ & ~+5.4~ & ~-1.5~ & ~+2.7~ \\
b) Inferred~~~         & ~-5.1~ & ~+9.0~ & ~-3.2~& ~+5.9~ & ~-1.9~ & ~+3.1~ \\
~~~c) Calculated~~    & ~-7.6~ & ~+11.0~ & ~-6.8~& ~+9.1~ & ~-5.4~& ~+6.5~  \\
\hline
\end{tabular}
\caption{TEM$_{n0}$ mode squeezing and anti-squeezing a) Corrected
for electronic noise. b) Inferred by taking detection and
propagation losses into account. c) Calculated taking the cavity
escape efficiency and relative pump power into account. }
\label{sqlamount}
\end{table}

The discrepancy between the calculated and the estimated
efficiencies, suggests that additional losses are present in the
system. A common extra possible loss factor is given by green
induced infra-red absorption (GRIIRA) \cite{Furukawa2001}. But
according to Furakawa \emph{et al.} no effect of GRIIRA should be
seen in our setup as we have chosen a $7\%$ MgO doped LiNbO$_{3}$
crystal. Nevertheless, we find that the losses seem to be induced by
the pump and are dependent on the H-G mode intensity distribution
and are therefore more important in an OPA pumped with a non-optimal
pump. Eq.~\ref{overlapcoeff10} and Eq.~\ref{overlapcoeff20} show
that the mode-overlap between the TEM{$_{00}$} pump and the
TEM{$_{10}$} and TEM{$_{20}$} seed are only 0.58 and 0.47,
respectively. Therefore more losses are induced and degrade the
measurable squeezing for the higher order H-G modes. This is also
what we find from the theoretical calculated efficiency,
Eq.~\ref{lossASQZ}. The comparison suggest that we have a decrease
in the cavity escape efficiency due to absorption in the crystal to
{$\eta_{cav, 00}$}$ = 0.76\pm0.02$, {$\eta_{cav, 10}$}$ =
0.62\pm0.02$ and {$\eta_{cav, 20}$}$ = 0.53\pm0.02$ for the
TEM{$_{00}$}, TEM{$_{10}$} and TEM{$_{20}$}, respectively. This
indicates that the absorption is getting larger for higher order H-G
modes due to the decrease in mode-overlap between the (non-optimal)
pump mode and the higher order H-G signal mode resonant in the OPA,
which is in good agreement with the simple mode-overlap model.

\begin{table}
\centering
  \begin{tabular}{|l|c|c|c|}
  \hline
        & ~TEM{$_{00}$~} & ~TEM{$_{10}$~} & ~TEM{$_{20}$~} \\ \hline
~~Est. efficiency~         & ~$0.79\pm0.04$~ & ~$0.76\pm0.04$~ & ~$0.73\pm0.04$~ \\
~~Cal. efficiency~        & ~$0.67\pm0.03$~ & ~$0.53\pm0.03$~ & ~$0.40\pm0.03$~ \\
\hline
\end{tabular}
\caption{Comparison between estimated and calculated detection
efficiency in our system.}\label{tablelos}
\end{table}

Losses are at this point limiting the amount of measurable squeezing
in the higher order H-G modes. However, we believe that there is a
potential for further improvement. A next experimental proposal
would be to generate optimal pump profiles in order to pump the OPA
more efficiently. As previously discussed, a way of achieving
perfect mode-matching between the TEM{$_{10}$} seed and the pump is
by creating a "multi-mode" pump beam consisting of a mixture of
TEM{$_{00}$} and TEM{$_{20}$}. For the  TEM{$_{20}$} mode, the pump
mode should be a mixture of TEM{$_{00}$}, TEM{$_{20}$} and
TEM{$_{40}$} modes. This multi-mode pump mode can be synthesize
using a spatial light modulator or by using a series of mode
transferring cavities. The later method for synthesize the
multi-mode is experimental very challenging.

\section{Conclusion}\label{conclusion}

In this contribution we demonstrate the generation of squeezed
light in the TEM$_{00}$, TEM$_{10}$ and TEM$_{20}$ modes. To our
knowledge this is the first demonstration of higher order
transverse mode squeezing using an OPA. Losses in the material
limit us presently to noise suppressions of 4~dB, -2.6~dB and
-1.5~dB. However, we can infer about -7~dB and -5~dB of noise
suppression inside the OPA. This is similar to the degree of
squeezing observed in conventional CW quantum optic experiments
and we believe that future improvements of our setup will allow us
to increase the observed amount of squeezing. We propose a method
to maximize the measurable squeezing of the present system using a
spatial light modulator to synthesize the multi-mode pump mode to
generate the optimal pump profiles for pumping the OPA more
efficiently and to minimize the nonlinear intra-cavity losses.

Using the transverse spatial properties of laser beams, we are now
able to produce all the required tools for higher order continuous
laser quantum optics experiments. The key components are the
ability to generate the H-G modes selectively with high efficiency
and the availability of simple and fully efficient modulation and
detection techniques. The way for parallel quantum information
processing with continuous variables in the transverse plane of a
laser beam using the basis of H-G modes is now open. In order to
test parallel quantum information protocols, we plan for future
experiments to produce spatial entanglement using two higher H-G
mode squeezers. Other types of experiments that could follow are
dense coding of spatial information, teleportation of spatial
information and spatial holography.

\section*{Acknowledgement}

We would like to thank Preben Buchhave, Jiri Janousek and Magnus Hsu
for many useful discussions. This work was supported by the
Australian Research Council Centre of Excellence scheme. ML is
supported by the Danish Technical Research Council (STVF Project No.
26-03-0304).

\bibliographystyle{unsrt}
\bibliography{MLv1}

\end{document}